\documentclass[review,3p,twocolumn]{elsarticle}

\usepackage[utf8]{inputenc}
\usepackage{tabularx}

\journal{Journal of Alloys and Compounds}

\begin{document}

\begin{frontmatter}

\title{Split of the magnetic and crystallographic states in Fe$_{1-x}$Rh$_{x}$Ge}

\author[PNPI,HPPI]{D.O. Skanchenko\corref{cor1}}
\cortext[cor1]{Corresponding author}
\ead{skanchenko\_do@pnpi.nrcki.ru}

\author[PNPI,SPSU,HPPI]{E.V. Altynbaev}

\author[HPPI]{V.A. Sidorov}

\author[LLB]{G. Chaboussant}

\author[LLB]{N. Martin}

\author[LPI]{A.E. Petrova}

\author[HPPI]{D.A. Salamatin}

\author[PNPI,SPSU]{S.V. Grigoriev}

\author[HPPI]{N.M. Chtchelkatchev}

\author[HPPI,LPI]{M.V. Magnitskaya}

\author[HPPI]{A.V. Tsvyaschenko}

\address[PNPI]{Petersburg Nuclear Physics Institute National Research Center "Kurchatov Institute", Gatchina, 188300 St. Petersburg, Russia}
\address[HPPI]{Vereshchagin Institute for High Pressure Physics, Russian Academy of Sciences, 142190 Troitsk, Moscow, Russia}
\address[SPSU]{Faculty of Physics, Saint-Petersburg State University, 198504 St. Petersburg, Russia}
\address[LPI]{Lebedev Physical Institute, RAS, 119991, Moscow, Russia}
\address[LLB]{Universite Paris-Saclay, CNRS, CEA, Laboratoire Leon Brillouin, 91191 Gif-sur-Yvette, France}

\begin{abstract}
We report on a comprehensive experimental and theoretical study of Fe$_{1-x}$Rh$_{x}$Ge compounds, within the entire concentration range $x \in \left[0.0 - 1.0\right]$, using X-Ray diffraction, small-angle neutron scattering, magnetometry and theoretical calculations. While FeGe and RhGe are single phase helimagnet and unconventional superconductor, respectively, an internal splitting of the crystallographic and magnetic states is found for intermediate compositions $x \in \left[0.2 - 0.9\right]$. A theoretical analysis of the stability of the two detected phases, together with the experimental data, indicate that this splitting preserves a common space group and occurs within single crystallites. Despite their apparent similarity, these two phases however display different magnetic structures, with distinct ferro- and helimagnetic character.
\end{abstract}


\end{frontmatter}

\section{Introduction}

The magnetic properties of compounds with a cubic B20 structure, such as {MnSi} and {FeGe}, are currently the subject of intense research. These compounds have a non-centrosymmetric crystallographic structure of B20 type, described by the space group {$P2_{1}3$}. The noncentrosymmetric arrangement of magnetic atoms in a B20-type lattice leads to the formation of a spin helicoid. It is known that this structure is based on a hierarchy of interactions between spins: ferromagnetic (FM) exchange interaction, antisymmetric Dzyaloshinskii–Moriya (DM) interaction, and anisotropic exchange interaction~
\cite{Dzyaloshinskii1964,Bak1980,Ishikawa1976,Lebech1989,Nakanishi1980}. The competition between FM ($J$) and DM ($D$) terms stabilizes the helical spin structure with the wave vector $k_{s} = D/J$.

The magnetic structure of binary compound {MnSi} undergoes a phase transition to an incommensurate 
helical spin state with a wave vector 
$\vec{k}_{s} = (2\pi/a)(\xi,\xi,\xi)$, where $\xi$ = 0.017, $T_C$ = 29.5 K. 
The spin helices in {MnSi} are oriented along four equivalent directions 
{$\langle 1,1,1 \rangle$}, so that in the absence of an 
external magnetic field, four types of magnetic domains 
are formed in the crystal \cite{Ishikawa1976, Ishikawa1977}.

Another compound similar to MnSi in terms of magnetic structure properties is {FeGe}~\cite{Lebech1989,Ludgren1970}. 
There are several significant differences between magnetic structures of {FeGe} and {MnSi}. 
The value of the temperature of magnetic ordering in case of {FeGe} is equal to $T_c$ = 278 K~\cite{Lebech1989}. This value is close to room temperature. 
Together with the possible applications of homochiral helical magnetic structure this amplify the researchers interest to transition metal monogermanides. 
Another difference between {FeGe} and {MnSi} is the magnitude of the wave vector of the magnetic 
helix in the ordered state, which in FeGe is directed along the {$\langle 1,1,1\rangle$} 
axes and equals to {$k \approx 0.09$ nm$^{-1}$} in compare to {$k \approx 0.36$ nm$^{-1}$} for {MnSi}. 
The value of the wave vector does not depend on the strength of the external field and 
depends weakly on temperature. It is known that, as the external magnetic field increases above 
the first critical field $H_{c1}$, the multidomain magnetic structure is rearranged 
into a single-domain conical helix with the wave vector $k_s$ directed along the field. 
With a further increase of the magnetic field up to $H_{c2}$, the conical spiral transforms 
into a field-induced ferromagnetic state~\cite{Ishikawa1976}. At temperatures near the $T_C$, 
in a certain range of fields $H_{c1} < H_{a1} \leq H \leq H_{a2} < H_{c2}$, the so called 
"skyrmion lattice" (or $A-phase$) appears. The $A-phase$ is an anomalous phase of the flip 
of the helix wave vector $k$ by 90$\circ$ with respect to the external magnetic field \cite{Muhlbauer2009}.

The definition of the magnetic chirality with respect to the structural chirality (left or right) is another intriguing feature of B20 compounds. 
It is well known that structural chirality "$\Gamma$" strictly determines the meaning 
of the magnetic chirality "$\gamma$" \cite{Tanaka1985,Ishida1985,Grigoriev2009,Grigoriev2010,Dyadkin2011,Grigoriev2013}. 
However, the relation between two chiralities is found to be different for various B20 compounds. 
It was fond that in case of Mn based compounds (such as MnSi and MnGe) the crystalline and magnetic 
chiralities have the same sign, while the chiralities are opposite to each other for Fe based ones, i.e., 
for FeGe. The study of the solid solution of {MnGe} and {FeGe} 
compounds with the help of small-angle neutron scattering reveals that the 
short-period helical structure of pure MnGe ($k = 2.3$ nm$^{-1}$) changes to a 
long-period helical structure of pure FeGe ($k = 0.09$ nm$^{-1}$) through a ferromagnetic-like 
transition at $x_c =$ 0.75 \cite{Grigoriev2013}. The possibility of the formation of ferromagnetic 
structure with non-zero value of $D$ was later theoretically explained by the competition between 
cubic anisotropy and DM interaction in solid solutions \cite{Sukhanov2015}. Similar result 
was later observed for Fe$_{1-x}$Co$_{x}$Ge solid solutions \cite{Grigoriev2014}. 
In both these families, the substitution involves solely 3d elements.

The substitution of 3d elements with 4d, such as {Rh}, leads to even more intriguing results \cite{Tsvyashchenko2016,Salamatin2021,Martin2017,Sidorov2018}. Measurements of the electrical resistance and magnetization of the binary compound of rhodium monogermanide {RhGe}, which is an isostructural analog of {MnSi} and {FeGe}, demonstrated a superconducting state below $T_c \sim$ 4.5 K~\cite{Tsvyashchenko2016}. The specific heat data confirmed the volume character of superconductivity in this compound. The replacement of {Mn} atoms with {Rh} brings the helical magnetic system to evolve unexpectedly with respect to Mn$_{1-x}$Fe$_{x}$Ge series \cite{Martin2017,Sidorov2018}. In particular, no flip of the magnetic chirality was observed and a split of the additional modulation of the period of magnetic order was found with SANS. The second reflection was explained as the presence of magnetic “twist grain boundary” phases, involving a dense short-range correlated network of magnetic screw dislocations with the cores described as nonradial double-core skyrmions \cite{Martin2017}.

Nevertheless, the binary compounds of Fe$_{1-x}$Rh$_{x}$Ge was not studied yet. The evolution of the magnetic structure of {FeGe} with {Rh} replacement of the magnetic atoms may lead to a nontrivial behaviour of the magnetic system and confirm the assumptions on the nature of helical ordering in B20 solid solutions. In particular, the nature of the magnetic system of {RhGe} and its possible applications is still to be revealed.

This work is devoted to the study of structural and magnetic properties of Fe$_{1-x}$Rh$_{x}$Ge 
compounds with $x$ = 0.0 -- 1.0. The evolution of physical characteristics with 
Rh replacement of Fe is followed. By means of theoretical and experimental methods. 
In particular, X-ray powder diffraction, small-angle neutron scattering (SANS) and 
magnetometry experiments have been performed. 
The split of the magnetic and crystallographic states in Fe$_{1-x}$Rh$_{x}$Ge was found. 
This result is qualitatively confirmed by the density functional theory.

\section{Experimental techniques}

For this work, Fe$_{1-x}$Rh$_{x}$Ge compounds were synthesized with $x$ from 0.0 to 1.0 with a step of 0.1. The samples were synthesized under a pressure of $P = 8$ GPa and $T$ = 1700 K in a high-pressure toroidal cell by melting a composition of Fe, Rh, and Ge in proportions corresponding to the desired composition of the solid solution~
\cite{Khvostantsev2004}. The samples are obtained in a polycrystal form with a crystallite size of about 10 {$\mu$m} (see details in~
\cite{Tsvyashchenko2012}). The crystal structure of the samples was studied by X-ray diffraction (XRD). The measurements were carried out at room temperature and atmospheric pressure using a Guinier camera – G670, Huber diffractometer (Cu K{$\alpha_{1}$}).

\begin{figure*}[h]
    \centering
	\includegraphics[width=1.0\textwidth]{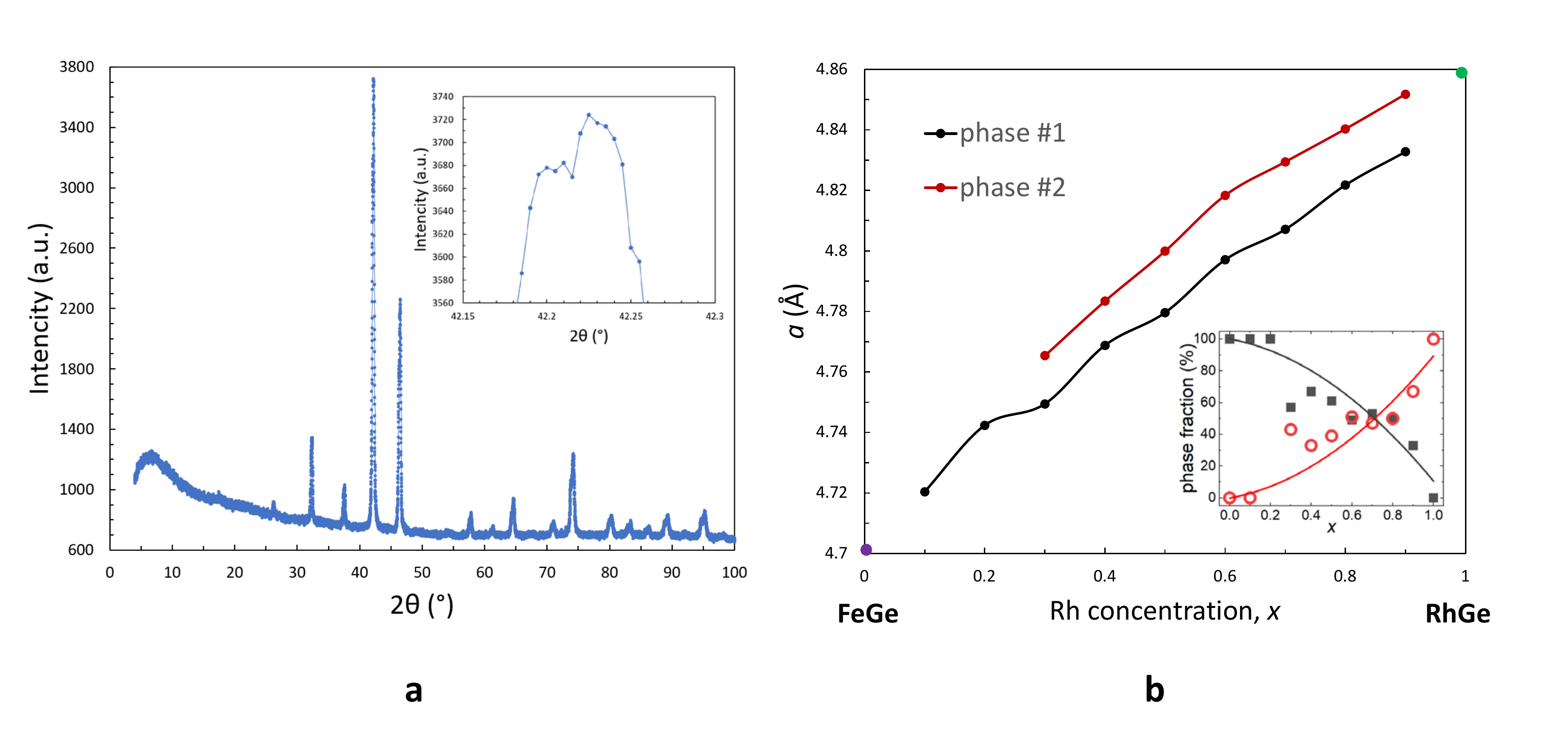}
	\caption{a: X-ray powder diffraction pattern of Fe$_{1-x}$Rh$_{x}$Ge at $x$ = 0.5;
b: Lattice constants versus Rh concentration in Fe$_{1-x}$Rh$_{x}$Ge. The green dot is the lattice constant for {RhGe}, the purple dot is the lattice constant for {FeGe}. The inset shows the mass fraction of each of the two phases as a function of the Rh concentration.
}
	\label{fig:1}
\end{figure*}

Magnetic susceptibility measurements have been performed in order to obtain the temperatures of the magnetic ordering for all synthesized compounds with ac method in a home-made coil system.

The magnetic structures were further probed using small-angle neutron scattering (SANS). The SANS experiments were carried out at the PA20 facility at the Laboratoire L\'eon Brillouin in Saclay, France. The neutron beam wavelength was chosen equal to 5 \r{A}, the sample-detector distance was varied in the range of 12–20 m in order to establish the {$Q$}-range available for investigation in the most optimal way for studying the magnetic structure of compounds. An external magnetic field was applied perpendicular to the incident neutron beam. Due to the high-pressure synthesis, the amount of sample for each composition was relatively small. In order to increase the amplitude of the signal, the studies were carried out as follows. Samples were first cooled to 5 K in a zero external magnetic field. After cooling the evolution of their magnetic structure was studied upon increasing the magnetic field. This made it possible to determine the various critical fields $H_{c1}$ and $H_{c2}$ at low temperatures for Fe$_{1-x}$Rh$_{x}$Ge compounds. Upon completion of the study of the field dependence of the magnetic structure, the magnetic field was removed. Further studies were carried out on the temperature dependence of the magnetic structure of the studied compounds. This procedure made it possible to determine $T_{C}$ with greater accuracy and unambiguously determine the presence or absence of helical magnetic ordering for all the studied compounds.

\section{Measurement results and theoretical analysis}

\paragraph{Crystal structure}
Investigating the crystal structure of the samples by X-ray diffraction, we found that all the Fe$_{1-x}$Rh$_{x}$Ge compounds crystallize into the B20 cubic structure. However, the diffraction peaks corresponding to the B20 structure split in the Rh concentration range from $x$ = 0.25 to $x$ = 0.9 (Fig.~\ref{fig:1}b).

This splitting can be described by two shifted peaks (Fig.~\ref{fig:1}a). In this case, no additional peaks were observed, which indicates the existence of identical isostructural phases with close cell parameters. The ratio of the intensities of the split peaks depends on $x$ pointing out the $x$-dependence of the volume fraction of each phase. The splitting of the peaks stays almost constant with $x$. X-ray patterns at $x$ = 0.0 - 1.0 can be fully described using a two-phase model with a B20 type structure (Fig.~\ref{fig:1}b). At $x$ = 0.0 - 0.2, as well as 1.0, only one phase is observed (see inset in Fig.~\ref{fig:1}b). At $x$ = 0.8, the fractions of the phases are equal. It should be noted that the lattice constants of the corresponding phases are extrapolated to the lattice constants for the binary compounds FeGe (at $x$ = 0.0) and RhGe (at $x$ = 1.0). Such stratification can indicate two different scenarios: either structural splitting occurs within one crystallite, i.e., the coexistence of two isostructural phases is observed, or it is a consequence of the crystallization of compounds with close values of Rh (Fe) concentrations, i.e. the existence of crystallites with different cell parameters is observed. However, the second scenario does not allow to explain the monotonic evolution of the volume fractions of different phases with $x$.

\paragraph{Ab initio simulations}

Our \textit{ab initio} computations were made within density functional theory (DFT). We used the projector--augmented--wave (PAW) pseudopotential method~
\cite{1t} as implemented in the VASP package~
\cite{Kresse1996}, with the PBE–GGA version~
\cite{Perdew1996} of the exchange-correlation potential. The calculations were converged with a plane-wave cutoff of 350 eV and a reciprocal-space resolution of $\sim 0.1$ \AA $^{-1}$ for \textbf{k}-point grids. The calculations of the partially disordered solid solutions Fe$_{1-x}$Rh$_{x}$Ge were done at $x$ = 1/4, 1/2, and 3/4, by replacing the due number of equivalent Fe atoms in the cubic B20 cell with Rh atoms. 

\begin{figure}[h]
    \centering
	\includegraphics[width=0.5\textwidth]{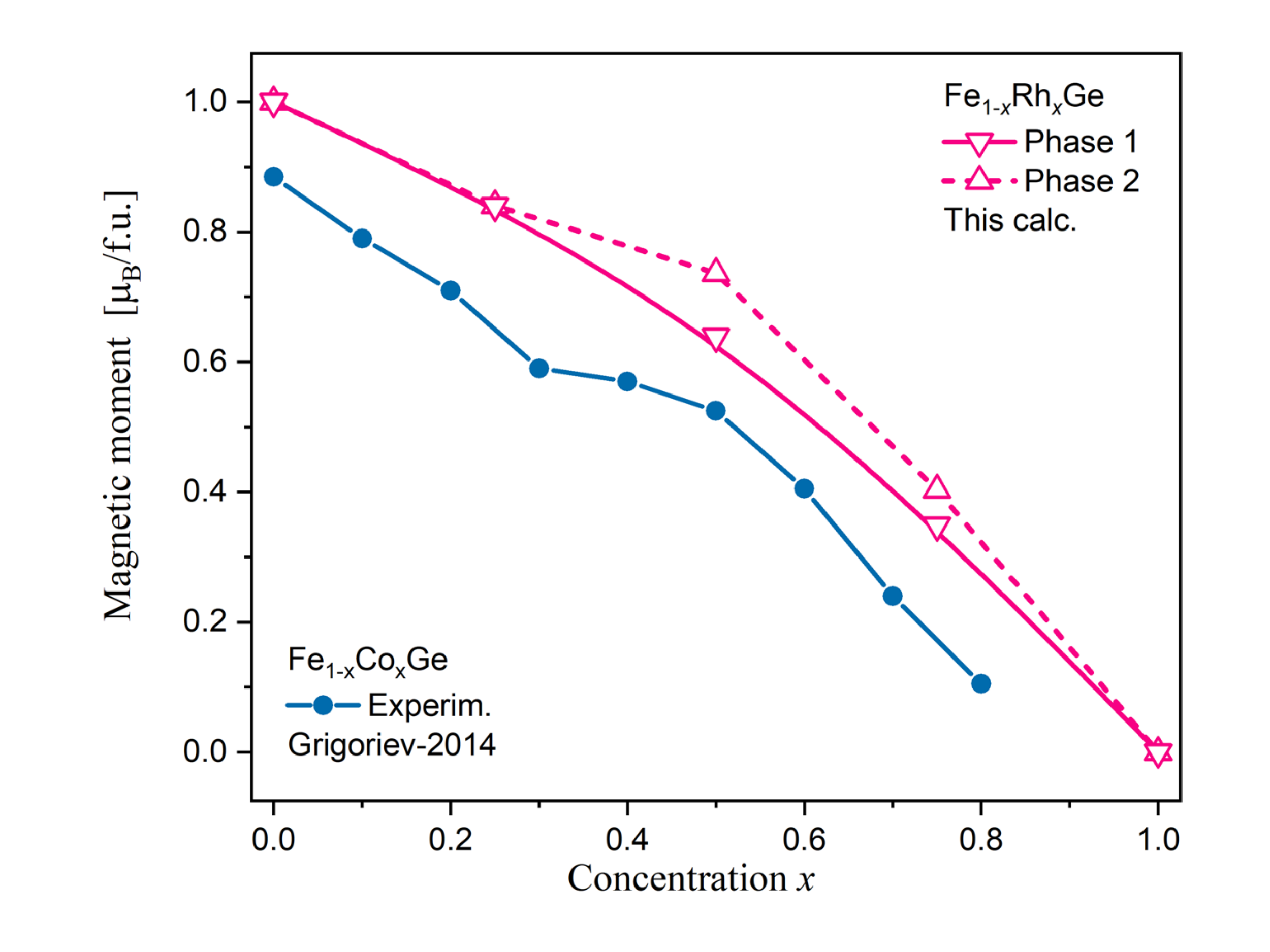}
	\caption{The calculated spin moments of phases 1 and 2 as functions of concentration $x$. (dashed pink curves). The experimental results~
	\cite{Grigoriev2014} for the isostructural and isovalent system Fe$_{1-x}$Co$_{x}$Ge are presented for comparison (full blue curve).
}
	\label{fig:2}
\end{figure}

It is worth noting that, due to the chemical inhomogeneity (replacement of Fe with Rh), the $P2_{1}3$ symmetry in the calculations is violated, thus formally the lattice becomes actually triclinic and, strictly speaking, does not belong to the B20 type. However, this symmetry-breaking distortion from B20 is quite small and only visible as a very weak broadening in our powder diffraction patterns (slightly more pronounced for phase 2). 

To test the hypothesis of two coexisting phases within one crystallite, we considered two crystal structures at each concentration $x$, in line with our experimental findings (see Fig.~\ref{fig:1}b). At $x$ = 0.5, the experimental lattice parameters were used ($a_1$ = 4.78~\AA~and $a_2$ = 4.80~\AA~for phases 1 and 2, respectively), while at $x$ = 0.25 and 0.75, interpolation between adjacent measured values was applied. At $x$ = 0.5 and 0.75, both phases 1 and 2 are found stable, with a slight ($<$ 0.17 eV/atom) energy preference for phase 1. At $x$ = 0.25, two phases turned out practically indistinguishable in enthalpy and other properties (recall that experimentally, phase 2 is only observed starting from $x$ = 0.3).

\begin{figure}[h]
    \centering
	\includegraphics[width=0.5\textwidth]{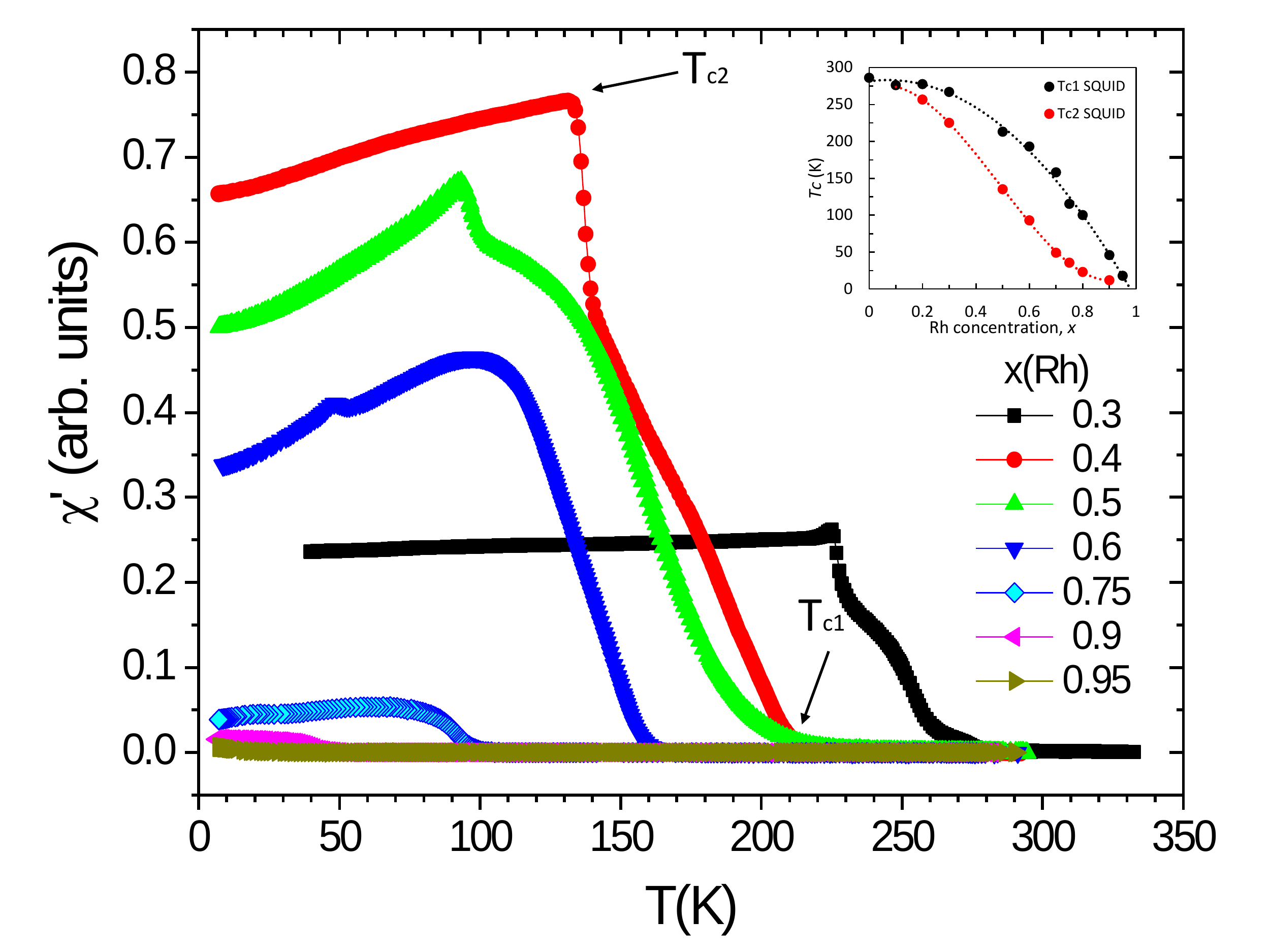}
	\caption{Temperature dependence of magnetic susceptibility for Fe$_{1-x}$Rh$_{x}$Ge.  The inset shows the concentration dependence of the critical temperatures in Fe$_{1-x}$Rh$_{x}$Ge.}
	\label{fig:4}
\end{figure}

Our spin-polarized calculations were performed for a collinear ferromagnetic (FM) alignment of spins. Here, this is a reasonable approximation, 
since the experimentally measured periods of magnetic structures in Fe$_{1-x}$Rh$_{x}$Ge (320--1400 \AA, see Fig.~\ref{fig:4}) 
are significantly longer than the unit-cell size ($\sim 5$ \r{A}). Thus, in calculations we do not distinguish a spin spiral and a 
FM state (infinite-period spiral). Fig.~\ref{fig:2} shows that the concentration dependence of evaluated magnetic moment per formula unit Fe$_{1-x}$Rh$_{x}$Ge 
is different for phases 1 and 2. For pure {FeGe}, the calculation yields a moment of $\sim 1$ $\mu${$_B$}/Fe, which compares well with the experiments and previous calculations~
\cite{Kamaeva2021,Wilhelm2011,Yamada2003,Neef2009}. The FM state in both structures is lower in energy (by $<$ 0.04 eV/atom) than the PM state. 

The cell magnetization is localized mostly at iron atoms and increases with their number. The magnetic moments at Ge atoms are small in magnitude ($<$ 0.07 $\mu${$_B$}) and antiparallel to the iron moments. The moments induced on the Rh sites are even smaller ($<$ 0.03 $\mu${$_B$}). Such magnetic arrangement is similar to that of Mn$_{1-x}$Rh$_{x}$Ge, which was calculated and confirmed by the XMCD measurements in our paper~
\cite{Sidorov2018}. For comparison, we also present in Fig.~\ref{fig:2} the experimental data~
\cite{Grigoriev2014} for the isostructural and isoelectronic system Fe$_{1-x}$Co$_{x}$Ge exhibiting analogous spin spiral. 

Both concentration dependences look very similar for FeGe doped with the isovalent 4d-Rh or 3d-Co. In our previous comparative study~
\cite{Chtchelkatchev2019} of {MnGe} doped with Rh/Co, a small systematic excess of the calculated magnetization values over the measured ones was ascribed to the noncollinearity of experimental magnetic arrangement. Phase 2 has a higher cell magnetization and iron moment. For example, the iron moment in FeRhGe$_{2}$ is equal to 1.37 $\mu${$_B$} and 1.60 $\mu${$_B$} for phases 1 and 2, respectively. The two phases also have different density of electronic states near the Fermi energy, which governs, in particular, the magnetic properties of metals.

\begin{figure}[h]
    \centering
	\includegraphics[width=0.5\textwidth]{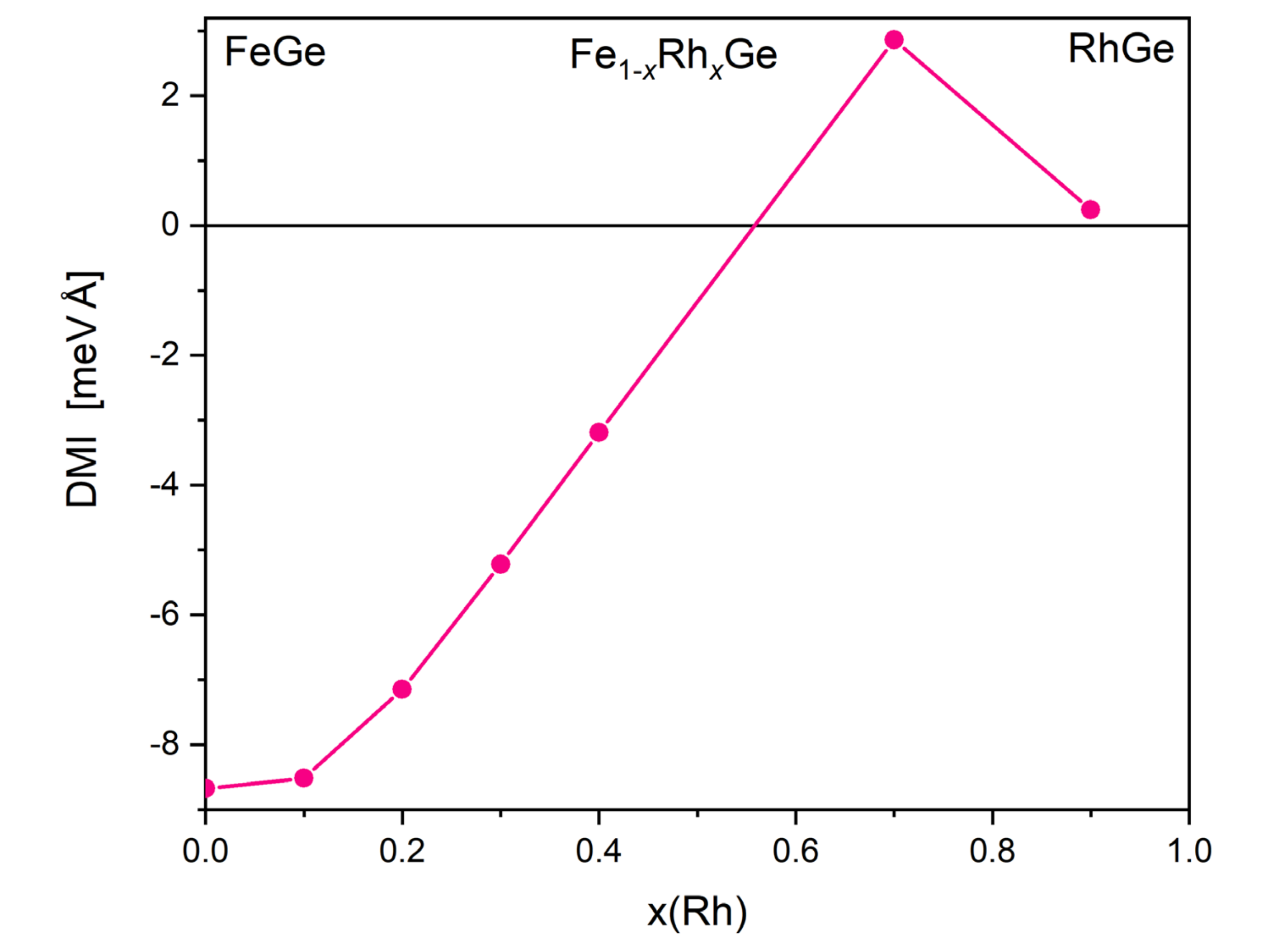}
	\caption{The estimated value of DM interaction in the Fe$_{1-x}$Rh$_{x}$Ge compounds.
}
	\label{fig:3}
\end{figure}

Since the experimental values of crystallographic parameters were used in our calculations, the resulting external pressures felt by the two phases are non-zero. Their difference is of about 2 GPa, with phase 2 being more non-equilibrium. Thus, qualitatively, we have a model situation of two coexisting isostructural phases with the same Rh content $x$ and different magnetic properties. This might indicate that the observed structural split occurs within a single crystallite and is not a consequence of the crystallization of compounds with close values of the Rh(Fe) concentration.

A real sample is evidently characterized by imperfections such as defects, chemical disorder and residual microstresses, typical of non-equilibrium phases synthesized under high pressure. As is known, parameters of the Dzyaloshinskii–Moriya interaction, which determine the presence and period of magnetic spiral, are highly sensitive even to very small mechanical deformations. The theoretical analysis and \textit{ab initio} calculations confirm this assumption. The theoretical analysis also a change in the sign of the DM interaction similar to that observed in Fe$_{1-x}$Mn$_{x}$Ge~
\cite{Koretsune2015}. This predicts the flip of the spin helix chirality in Fe$_{1-x}$Rh$_{x}$Ge with $x$ increase at $0.5 \leq x \leq 0.6$ (Fig.~\ref{fig:3}). The relatively small value of calculated DM constant in the region of $x > 0.6$ indicates a possible ferromagnetic structure of RhGe-based compounds according to~\cite{Sukhanov2015}.

\paragraph{Magnetic properties}

In order to verify the theoretical predictions the magnetometry and SANS measurements were performed. Fig.~\ref{fig:4} shows the results of magnetometry and the temperature dependencies of the magnetic susceptibility ($\chi$) of Fe$_{1-x}$Rh$_{x}$Ge compounds with different Rh concentrations.

As can be seen from the obtained experimental data it is possible to determine two critical temperatures of the magnetic transition for samples with $x > 0.2$, which are defined as the point of intersection with zero of the magnetic susceptibility, $T_{c1}$, and as the temperature at which the susceptibility is maximum, $T_{c2}$. This points out two possible phase transitions to different magnetic states. The $x$ dependence of the critical temperatures $T_{c1}$ and $T_{c2}$, obtained as a result of the analysis of experimental magnetometry data, is shown in the inset in Fig.~\ref{fig:4} (black and red dots ).

\begin{figure*}[h]
    \centering
	\includegraphics[width=1\textwidth]{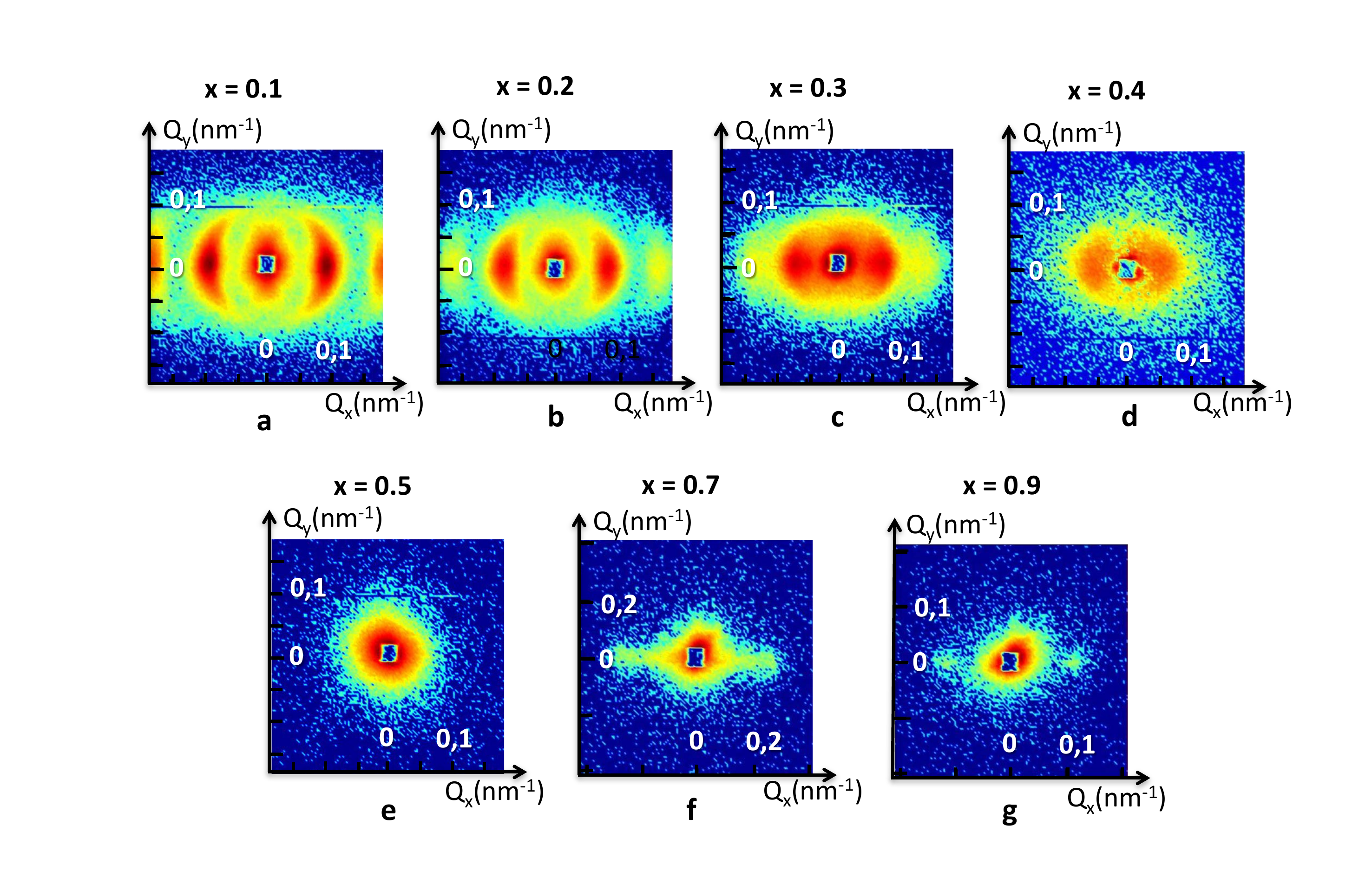}
	\caption{SANS intensity maps obtained on Fe$_{1-x}$Rh$_{x}$Ge samples, obtained at a temperature of 5 K in a zero external magnetic field after the application of an external magnetic field.
}
	\label{fig:5}
\end{figure*}

SANS measurements were carried out in order to determine the type of the magnetic structure of the compounds. The SANS patterns obtained for the samples of Fe$_{1-x}$Rh$_{x}$Ge at $T = 5$ K in a zero external magnetic field after the application of an external magnetic field above $H_{C2}$ are presented in Fig.~\ref{fig:5}. As it is shown in Fig.~\ref{fig:5}a two Bragg reflections are observed on both sides of the direct beam for Fe$_{0.9}$Rh$_{0.1}$Ge, which corresponds to a single-domain helical structure. With $x$ increase, the reflections shift towards each other, which indicates a decrease of the helix wave-vector value (Fig.~\ref{fig:5}a-d). A further increase of Rh concentration leads to the disappearance of the signal from helical structure and only diffuse scattering remains (Fig.~\ref{fig:5}e). With further substitution of Fe atoms by Rh up to $x$ = 0.7, the intensity peaks appear again (Fig.~\ref{fig:5}f). Further $x$ increase leads to a decrease of the helical wave-vector value (Fig.~\ref{fig:5}g).

\begin{figure}[h]
    \centering
	\includegraphics[width=0.5\textwidth]{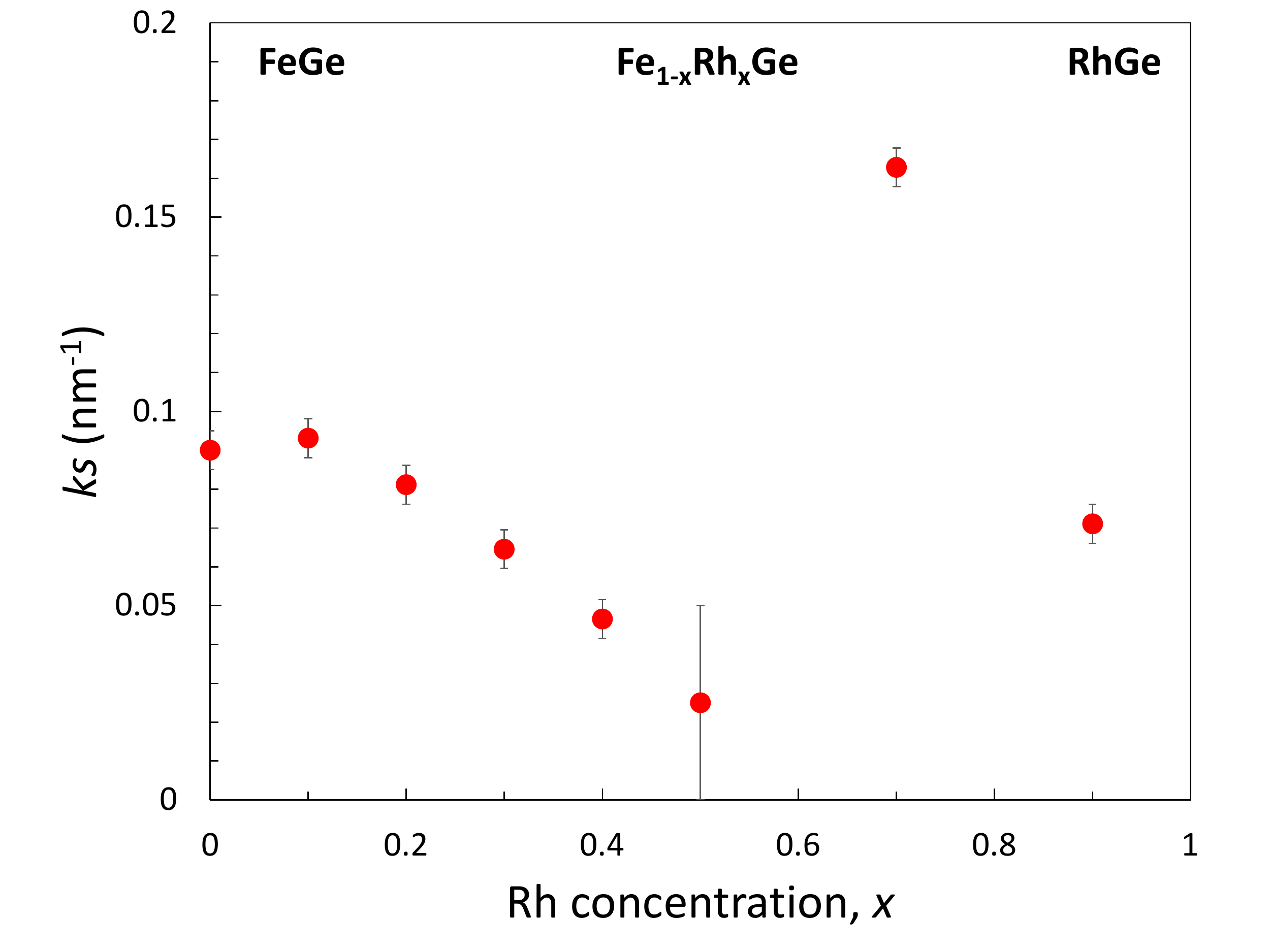}
	\caption{Dependence of the helix wave vector $k_{s}$ on the concentration $x$
at $T$ = 5 K
}
	\label{fig:6}
\end{figure}

The neutron scattering intensity was averaged over the azimuthal angle. The value of the helical wave vector $k_{s}$ was determined from the neutron scattering profile as the value of the momentum transfer corresponding to the maximum neutron scattering intensity. Fig.~\ref{fig:6} shows the $x$-dependence of the helix wave-vector value $k_{s}$. It can be seen that at concentrations 0 $\leq x \leq$ 0.4, the wave-vector value decreases with $x$ from $k_{s} = 0.09\pm0.005$ {nm}$^{-1}$ at {$x$} = 0, i.e., for pure {FeGe}~
\cite{Grigoriev2013}. For compounds with $x$ = 0.5 and 0.6, the scattering from the helical structure was not observed within the SANS experiment. This indicates the transition of the studied magnetic system to the ferromagnetic state. Since the wave-vector value $k_{s}$ is small and evolves monotonically with $x$ within the range 0 $\leq x \leq$ 0.4, it can be argued that the helical structure of these compounds is satisfactorily described by the model developed for {FeGe}~
\cite{Bak1980}. A similar vanishing of the wave vector for intermediate concentrations was found earlier for Fe$_{1-x}$Mn$_{x}$Ge and Fe$_{1-x}$Co$_{x}$Ge compounds~
\cite{Grigoriev2013,Grigoriev2014}. This phenomenon was interpreted as a flip of the magnetic chirality with  $x$ and later confirmed by calculations~
\cite{Koretsune2015,Chen2015,Kikuchi2016}. Analysis of the DM constant for the series of compounds studied in this work also satisfactorily describes the behaviour of the wave vector of the magnetic helix over the entire range of concentrations. In addition, the sign change of the DM constant in the concentration range $x$ = 0.5 – 0.6 (Fig.~\ref{fig:3}) suggests the reversal of the magnetic chirality in these compounds. A further increase in the Rh concentration leads to a sharp increase, for $x$ = 0.7, and then again a decrease, for $x$ = 0.9, of the wave-vector value (Fig.~\ref{fig:6}), which corresponds to the theoretical estimation of the DM constant (Fig.~\ref{fig:3}).

\begin{figure}[h]
    \centering
	\includegraphics[width=0.5\textwidth]{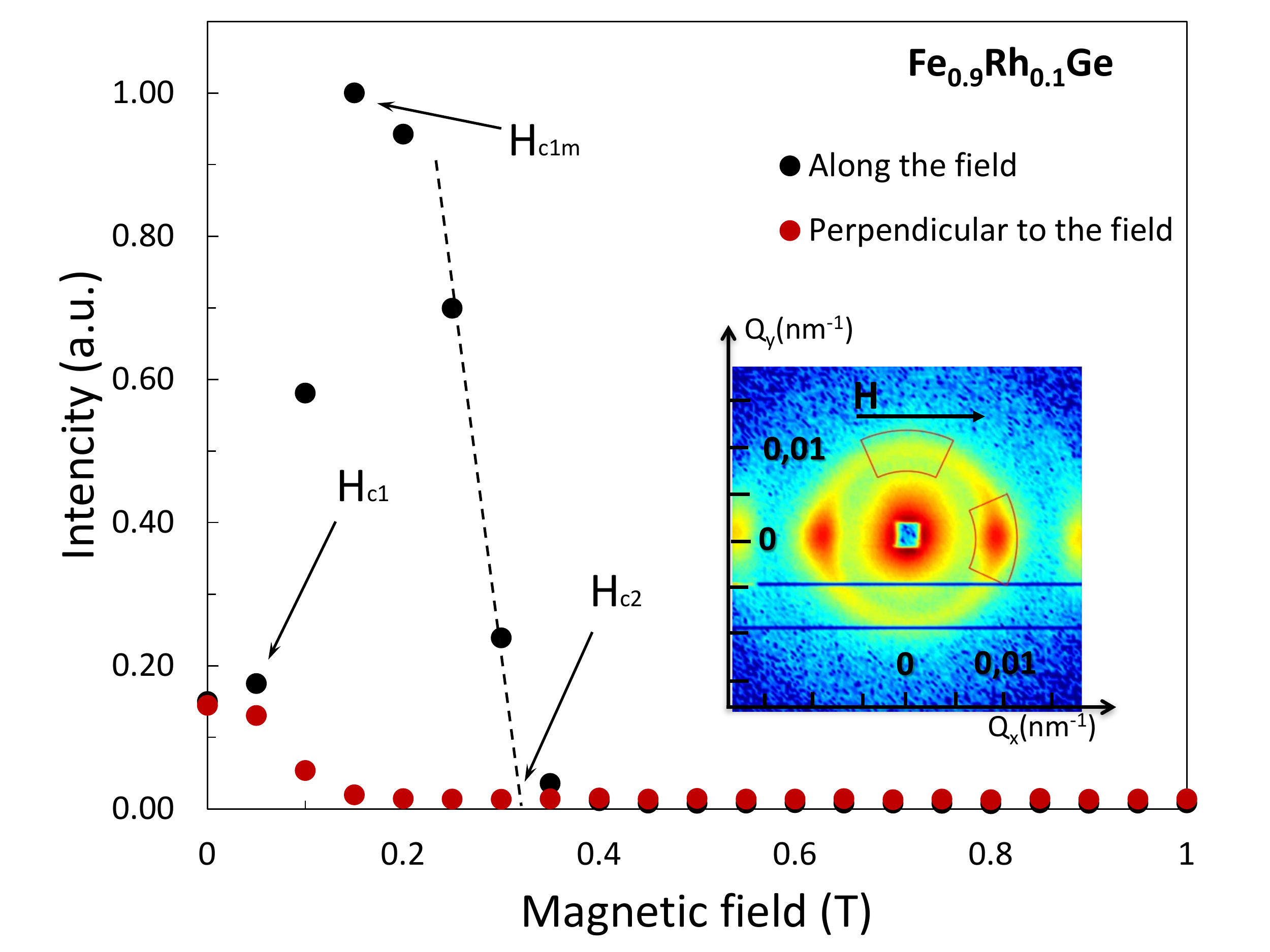}
	\caption{Evolution of the magnetic structure of Fe$_{0.9}$Rh$_{0.1}$Ge with increasing external magnetic field at $T =$ 5 K. An example of integration is shown using the red sector on the SANS map obtained at $T =$ 5 K, in the inset.
}
	\label{fig:7}
\end{figure}

The neutron scattering intensity also varies with the magnetic field. To analyze the corresponding changes in the magnetic structure, the scattering intensity was integrated in the sectors shown in the inset of Fig.~\ref{fig:7}. The $H$-dependence of the intensity on the external field, obtained as a result of integration at $T = 5$ K after zero-field cooling, is shown in Fig.~\ref{fig:7}. Based on the field dependence of the scattering intensity, three critical values of the magnetic field for phase transitions can be determined: $H_{c1}$, $H_{c1m}$ and $H_{c2}$ . The critical field $H_{c1}$, at which the intensity of the Bragg reflection starts to increase, indicates the beginning of the process of reorientation of the magnetic spirals in the direction along the external magnetic field. This implies a transition from a multidomain state to a single domain conical structure. The field $H_{c1m}$, at which the intensity of neutron scattering along the external magnetic field reaches its maximum, marks the end of the reorientation process and the transition of each grain in the polycristalline sample to the single-domain conical state. The critical field $H_{c2}$, determined by extrapolating to zero the linear decay of the longitudinal scattering intensity with increasing external magnetic field, marks the transition to the field-induced ferromagnetic state.

\begin{figure}[h]
    \centering
	\includegraphics[width=0.5\textwidth]{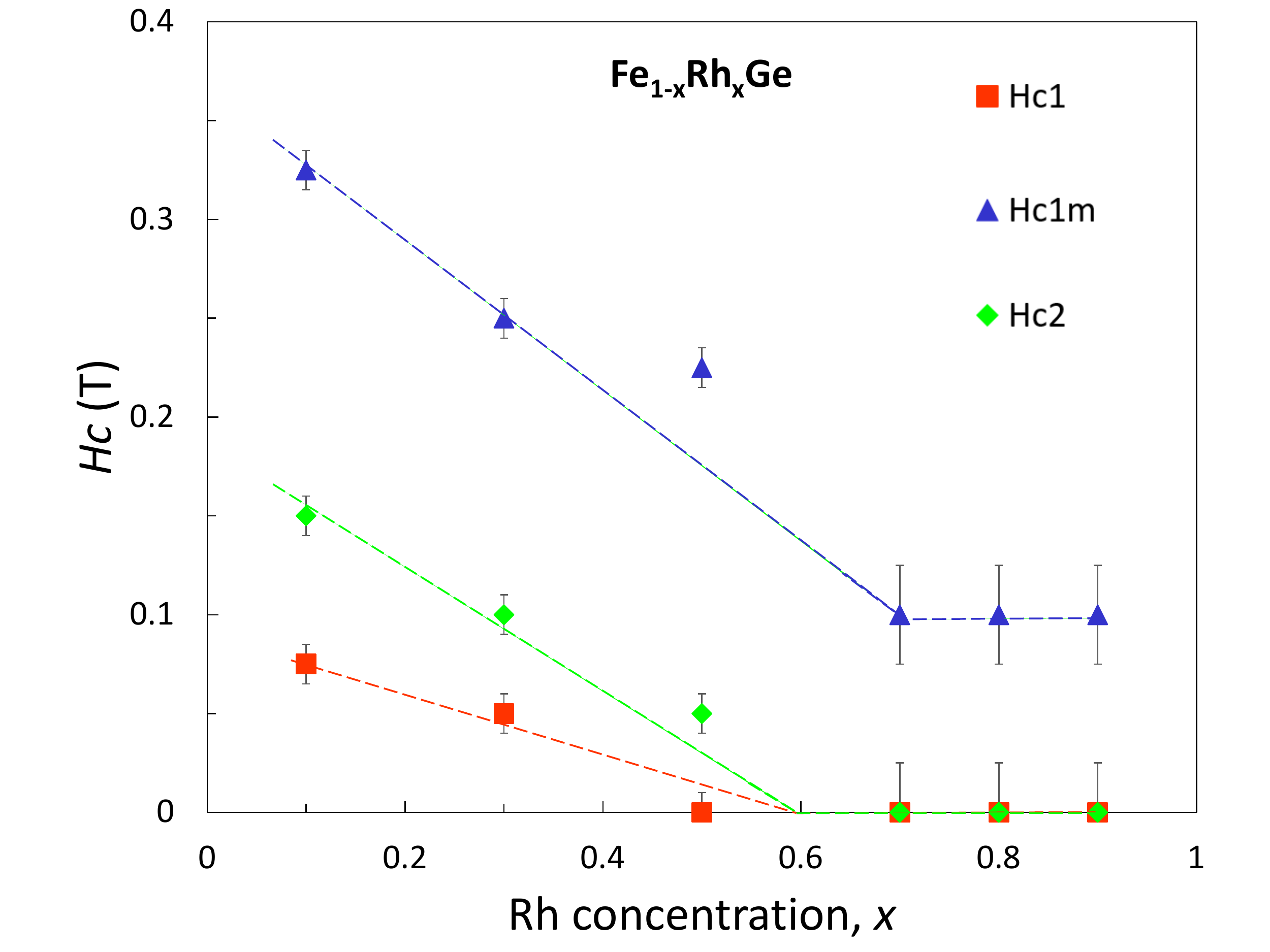}
	\caption{Dependence of the critical fields $H_{c1}$, $H_{c1m}$ and $H_{c2}$ on the concentration $x$
}
	\label{fig:8}
\end{figure}

The $x$-dependence of the critical fields $H_{c1}$, $H_{c1m}$ and $H_{c2}$ obtained for all the studied compounds at $T$ = 5 K is shown in Fig.~\ref{fig:8}. It can be seen in Fig.~\ref{fig:8} that an increase of $x$ up to 0.5 leads to a decrease in all critical fields. With a further increase in the Rh concentration, $x >$ 0.5, no change in the critical fields was found. This might be connected to the weak signal of helical structure from samples with $x \leq 0.7$ after zero field cooling and low volume fraction of helical structure in the compound (inset of Fig.~\ref{fig:1}b).

\begin{figure}[h]
    \centering
	\includegraphics[width=0.5\textwidth]{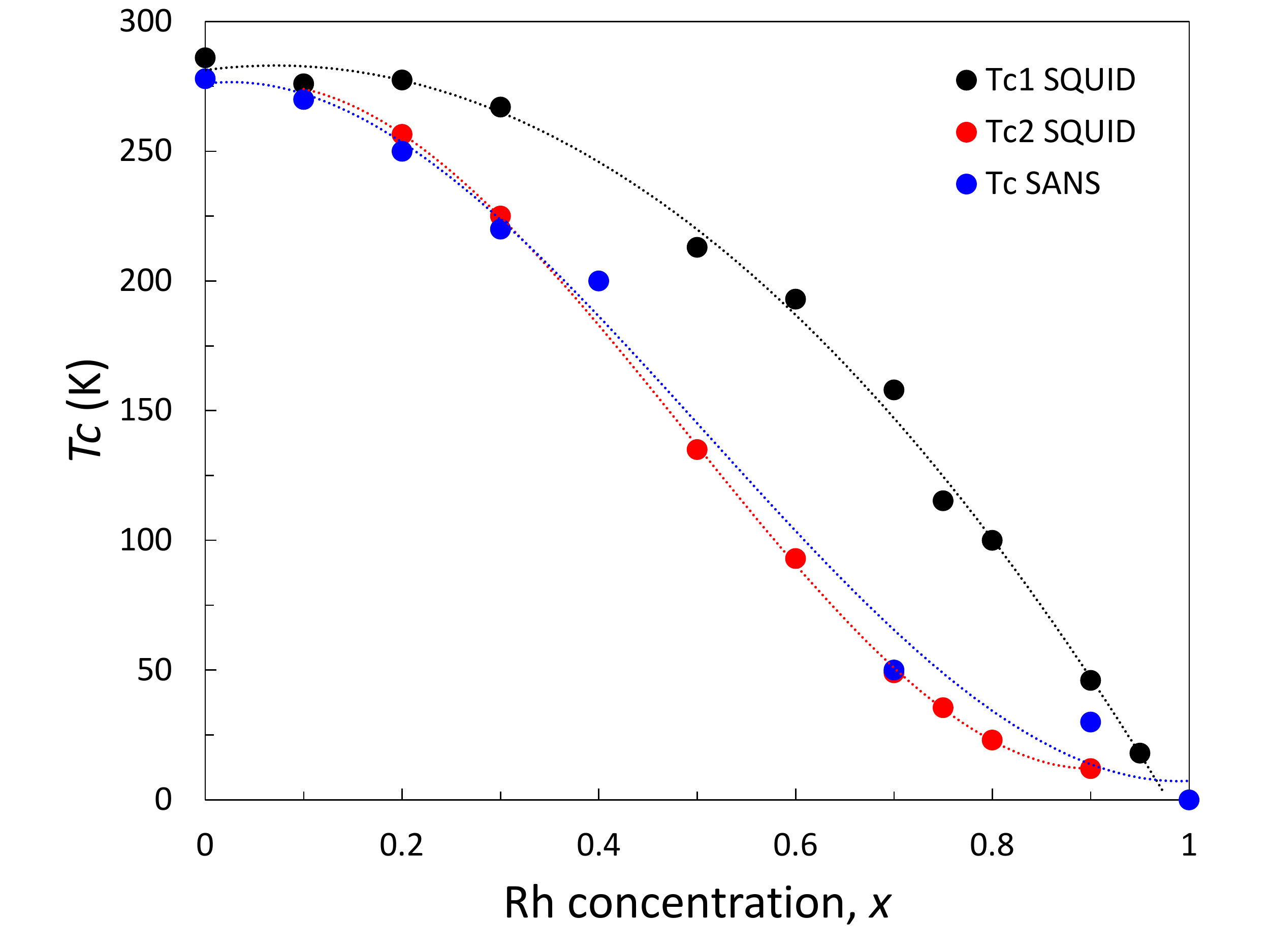}
	\caption{Concentration dependence of critical temperatures in Fe$_{1-x}$Rh$_{x}$Ge. Black and red dots - magnetometry; blue dots - SANS
}
	\label{fig:9}
\end{figure}

In order to determine the critical temperature of the transition from the paramagnetic state to the helical state, the dependence of the integral intensity of the Bragg peak on temperature was measured for all samples. The temperature $T_c$ corresponds to the appearance of magnetic Bragg peaks and to a sharp increase in scattered intensity upon cooling. The obtained dependence of the critical temperature $T_c$ on $x$ is plotted in Fig.~\ref{fig:9}. It can be seen that the ordering temperature $T_c$ decreases with increasing concentration $x$ from 278 K for {FeGe}~
\cite{Grigoriev2013} to zero for {RhGe}. 

This concentration dependence also clearly shows that the curve obtained from the analysis of the experimental SANS data corresponds to the curve obtained from the magnetometry data. Hence, we can assume that the higher transition temperature obtained with magnetometry corresponds to the ferromagnetic state, while the lower critical temperature corresponds to the helimagnetic transition.

Comparison of the data with theoretical analysis suggests that two magnetic phase states in the sample correspond to different isostructural crystalline phases. In addition, taking into account that the {FeGe} compound exhibits helical properties at low temperatures and has a smaller cell parameter compared to {RhGe}, we assume that the helical magnetic structure of the crystalline phase corresponds to a smaller cell parameter in the concentration range $x \in \left[0.2 - 0.9\right]$. This is also confirmed by the disappearance of the helical magnetic order with an increase of the Rh concentration to a value of 1.0, which posses weak ferromagnetic properties at low temperatures that might be explained by the decay of DM constant in compare to cubic anisotropy in RhGe-based compounds~\cite{Sukhanov2015}.

\section{Conclusions}

Summarizing the experimental and theoretical data, we can conclude that we observe an internal splitting of the crystal structure of Fe$_{1-x}$Rh$_{x}$Ge compounds in a wide range of concentrations $x \in \left[0.2 - 0.9\right]$. A theoretical analysis of the stability of the two detected isostructural phases indicates that structural splitting occurs within a single crystallite and is not a consequence of the crystallization of compounds with similar Rh(Fe) concentrations.

Another result of the theoretical analysis is the difference of the magnetic moments for two different 
crystallographic phases. This indicates the possible coexistence of 
two different types of magnetic order in the sample. 
The magnetisation measurements made for Fe$_{1-x}$Rh$_{x}$Ge reveal 
two magnetic phase transitions with temperature in a wide range of concentrations $x \in \left[0.2 - 0.9\right]$. 
Therefore we can conclude that that two magnetic states with different $T_C$ were observed: 
ferromagnetic state with a high-temperature phase transition and 
helimagnetic state with a phase transition occurring at lower temperatures. 
The helical nature of the magnetic order at low temperatures was independently confirmed by SANS. 
Similarly to Fe$_{1-x}$Co$_{x}$Si and  Fe$_{1-x}$Mn$_{x}$Ge compounds the flip of the magnetic chirality 
was observed for helical magnetic structure in Fe$_{1-x}$Rh$_{x}$Ge at $x \sim 0.6$, 
which is confirmed by theoretical calculations of DM constant in the compound. 
This also means that the crystallographic and magnetic chiralities coninside in case of 
{RhGe}-based compounds as well as for {MnSi} or {MnGe}-based ones. 
As long as {FeGe} posses the helimagnetic ordering at low temperatures, 
it can be concluded that the helical structure correspond to a 
crysatllographic phase with smaller cell parameter and the isostructural phase 
with a larger cell parameter posses ferromagnetic ordering within the $x$-range between 0.2 and 0.9. 
Also the assumption could be made that the {RhGe}-based compounds posses the 
ferromagnetic behaviour despite the lack of the inversion symmetry and nonzero value of DMI in the compounds. 
The monotonic behaviour of volume fraction of each of isostructural crystallographyc phases confirms this assumption.

The ferromagnetic nature of the {RhGe}-based compounds could be explained by the predictions made theoretically in \cite{Sukhanov2015}. That means that one should expect a much stronger influence of the cubic anisotropy on the formation of the magnetic structure of {RhGe} in compare to {FeGe}. Possibly the differences in electronic structure of the compounds with 3d or 4d magnetic atoms are responsible for such a result. Therefore, further investigation of replacement of {Mn} or {Fe} atoms with 4d elements in B20 binary compounds are important for development of accurate theoretical model of these magnetic systems.

\section{Acknowledgment}

This work was supported by Russian Science Foundation under Grant No. 22-12-00008. The numerical calculations were performed using computing resources of the federal collective usage center Complex for Simulation and Data Processing for Mega-science Facilities at NRC "Kurchatov Institute" (http://ckp.nrcki.ru/) and supercomputers at Joint Supercomputer Center of RAS (JSCC RAS). Part of calculations was carried out on the "Govorun" supercomputer of the Multifunctional Information and Computing Complex, LIT JINR (Dubna).

\bibliography{FeRhGe}

\end{document}